\documentstyle[epsfig]{aipproc}
\begin{document}
\title{The Extent of the Spectral Bias in BATSE: The True Distribution of the
$\nu F_{\nu}$ Peak Energy}

\author{Nicole M. Lloyd and Vah\'e Petrosian}
\address{Center for Space Science and Astrophysics \\ Stanford University\\
Stanford, California 94305}

\maketitle

\begin{abstract}
The distributions of spectral characteristics and their correlations
with fluence, peak flux, or duration, are essential in understanding the nature of GRBs.
However, the selection effects involved in detecting GRBs can distort these
distributions.  Here, we discuss how to deal with selection effects involving 
the peak energy $E_{p}$ of the GRB $\nu F_{\nu}$ spectrum, which suffers from both
an upper and lower threshold. We describe a new method to
account for this double-sided
truncation, and show that the true distribution of $E_{p}$ is significantly 
different from the observed distribution.
\end{abstract}

\section*{Introduction}

 The spectral properties of GRBs provide the most direct information about the 
 physical processes associated with the event. In particular, the distribution of the spectral
 parameters and the correlation between these and other GRB characteristics can shed 
 significant light on the radiation mechanisms and energy production of the burst. 

Several studies have obtained the spectral parameters for the brightest GRBs (1) 
and investigated correlations between these
parameters and peak flux (2),
duration (3) and
spatial distributions (4).

 However, as pointed out by Piran and Narayan (5),
caution is required in obtaining these distributions and correlations. 
Several selection effects come into play  
in detecting GRBs, which limit the information we can obtain from the data. 
These selection criteria often
truncate the data, which - if not accounted for properly -  can result in 
incorrect distributions and correlations.

In some cases, such as with $Log(N)-Log(S)$ distributions, the data is truncated
from only one side; methods dealing with this kind of truncation 
were discussed by Petrosian (6) and Efron and Petrosian (7)
and applied to GRBs by Lee and Petrosian (8).  

However, in other cases 
the data truncation can be more complicated - some parameters can have both a lower $and$
upper
limit.  
A good example of this is the peak energy $E_{p}$ of the $\nu F_{\nu}$ spectrum;
this is the focus of our paper.

\section*{Spectral Parameters}

It is well known that the $\nu F_{\nu}$ spectrum of a GRB peaks at an energy $E_{p}$ with
    power law indices of $\alpha$ and $\beta$ below and above this energy, respectively.  Band
   et al. have presented a useful, smooth form of such a spectrum:

\begin{equation}\label{equation}    F(E) =
      \cases{ A/E_{p} (E/E_{p})^{\alpha} exp[(\beta-\alpha)(E/E_{p}-1)],&    $ E < E_{p}$ \cr
       A/E_{p} (E/E_{p})^{\beta},& 
       $E >  E_{p}$}
       \end{equation}
$F(E)$
can be the photon or energy fluence or flux. In this paper, we will deal with the energy fluence
in which case $E F(E) \propto \nu F(\nu)$,  and A is in units of 
ergs/cm$^{2}$.  
Our aim is to demonstrate how to obtain bias
free distributions for $E_{p}$.  

\section*{Data}

 Determining accurate values for the spectral parameters of a large number
   of GRBs is difficult.  We use the ratio of
   the fluences in the four channel BATSE LAD detectors to determine $\alpha$, $E_{p}$, and $A$
   via the downhill simplex method. The index $\beta$ was derived from an assumed gaussian distribution;
   the form of this distribution was inconsequential to the calculations as a whole. 

     Given the values of $\alpha,\beta$, $A$, and the fluence thresholds for each
     burst,  
     we may ask (in the spirit of the $V/V_{max}$ test):
     \sl What $E_{p}$ brings the fluence below the
     threshold value? \rm
     [See Petrosian and Lee (9) for
        how the threshold is obtained from the C$_{max}$ and C$_{min}$ values in the catalog.]
       It can be shown, as a result of the BATSE trigger condition,  that there is both an upper limit $E_{p_{max}}$
     and lower limit  $E_{p_{min}}$ (another way of stating this is that BATSE is most sensitive to bursts with $E_{p}$ in
     the 50-300 keV trigger range).
     As pointed out by Piran and Narayan (5), 
     this
     introduces some bias in the distribution of $E_{p}$.  In order to quantify the severity of
     this bias, we have determined the values of $E_{p_{min}}$ and $E_{p_{max}}$ for 433 bursts in
     the 3B catalog (See 
     Figure 1);
     \begin{figure}
     \centerline{\epsfig{file=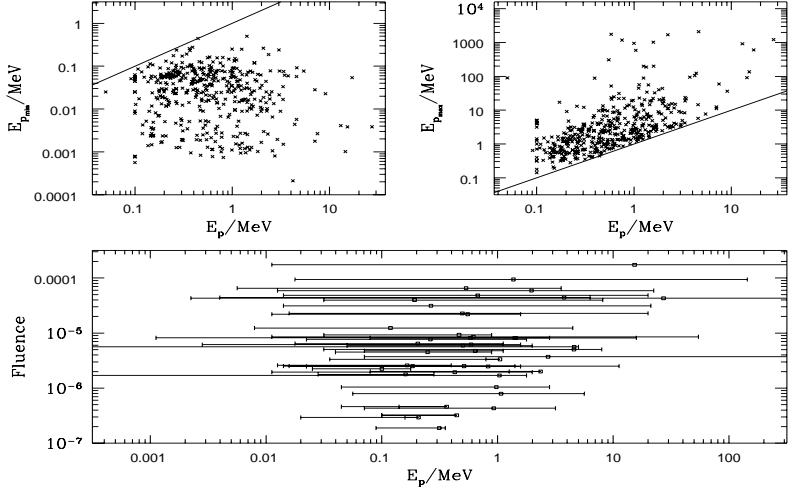, height = 7cm,width = 11cm,angle=270}}
     \caption{$E_{p}$ vs. $E_{p{min}}$ (top left) and $E_{p}$ vs.
$E_{p_{max}}$ (top right); the solid line shows the truncation boundaries.  The lower plot shows Fluence vs.
$E_{p}$,  displaying the upper and lower limits on $E_{p}$ for a representative number of bursts.}
     \end{figure}
     from this, we obtain the corrected distribution of $E_{p}$ and compare it with the
     raw distribution.  

\section*{Method}

Our aim is to correct the observed distribution given that the observations can detect only
bursts with $E_{p}$ limited to the interval
$[E_{p_{min}},E_{p_{max}}]$.
Recently, in collaboration with B. Efron, we have developed a method
     to deal with this kind of data  truncation ( 
     this
     method is a generalization of the one sided monotonic truncation case employed in
     our earlier studies (6), (7)).

  Consider data points $(x_{i},y_{i})$ (or in our case $(E_{p},F))$,
 where $x_{i}$
 has lower and upper limits  $l_{i}$ and $u_{i}$, respectively;  
 $x_{i}  \in T_{i} = [l_{i},u_{i}]$.
Let $f(x)$ be the true distribution of $x$, which
would be observed if there were no truncations. 
However, because $x_{i}$ is limited to $T_{i}$, we observe the
conditional distribution $f(x_{i}|T_{i}) = f(x_{i})/F(T_{i})$ where 
$F(T_{i}) = (\sum_{j} f(x_{j})$ : $x_{j} \in T_{i}$) 
is the probabilitly that $x$ exists in
$T_{i}$. We define 

\hspace{1.75in} $\bullet$ $f_{i} = f(x_{i})$, 

\hspace{1.75in} $\bullet$ $F_{i} = F(T_{i})$, and

\hspace{1.75in} $\bullet$ $J_{i,j} = \cases{1,&  $x_{j} \in T_{i}$\cr
                   0,&  $x_{j} \notin T_{i}.$}
$ \\
The goal is to estimate $f(x)$ from $l_{i},u_{i}$, and $x_{i}$
                assuming all $N$ cases are
                independently distributed.  
The procedure for this amounts to solving the following three equations iteratively:
      
\begin{enumerate}
\item{                 $F_{i} = J_{i,j} f_{j}$ \hspace{1.25in}(definition)},
\item{                 $1/f_{i} = J_{j,i} F_{j} + constant $ \hspace{0.2in} (maximum likelihood assertion)},
\item{                 $\sum_{i=1}^{N} f_{i} = 1$\hspace{1.25in}(normalization)}.
\end{enumerate}
Convergence is reached when $constant$ goes to zero.

  This method can be used to determine univariate cumulative and differential
    distributions, as well as correlations between
    relevant variables (e.g. fluence and $E_{p}$ or peak flux and $E_{p}$), (10).  

\section*{Results}

In Figure 2, we present the uncorrected and corrected cummulative and differential
     distributions of $E_{p}$. Note the large number of $E_{p}$'s
     \begin{figure}
\centerline{     \epsfig{file=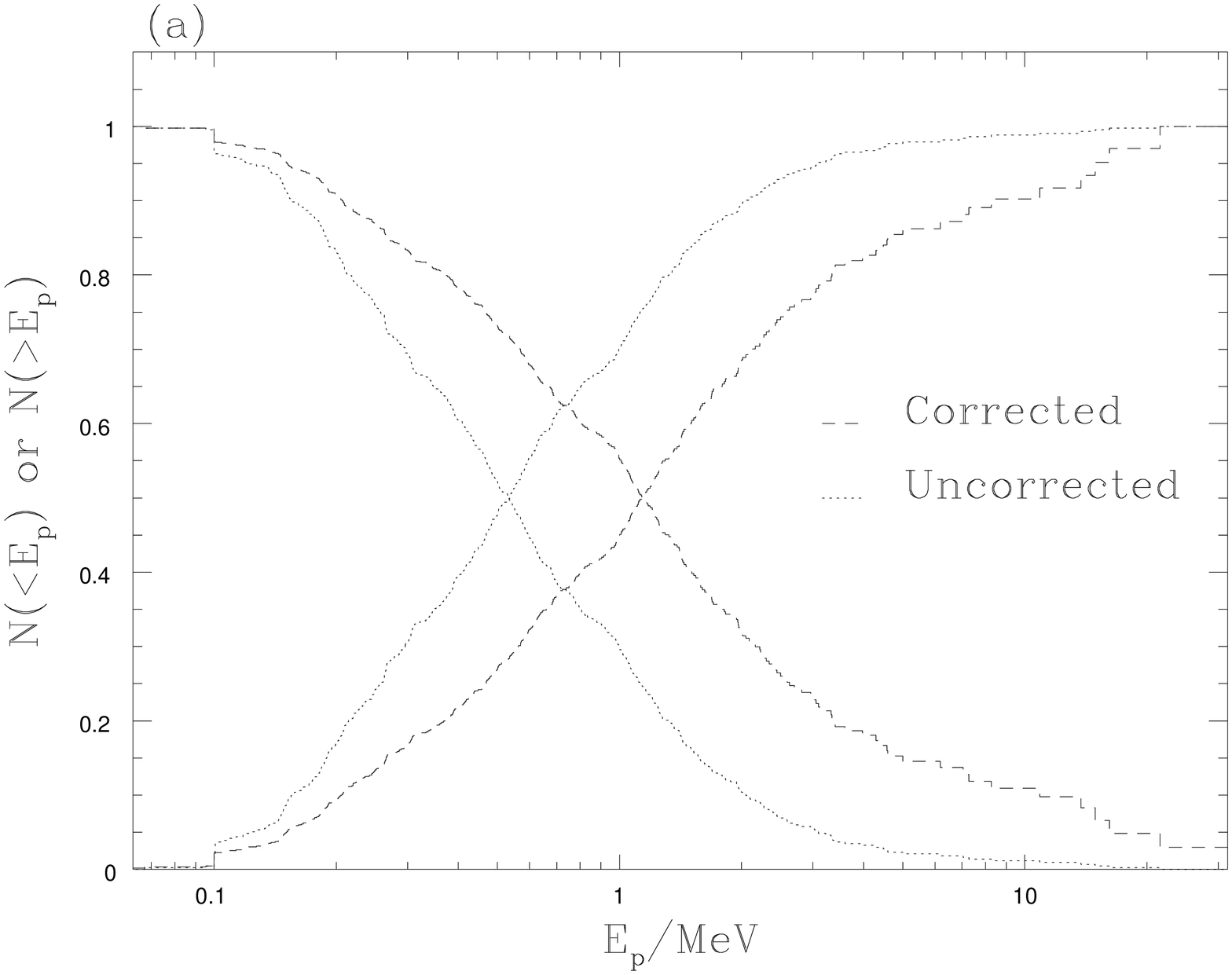,width=6.8cm,height=6cm,angle=270}
 \label{(a)}
     \epsfig{file=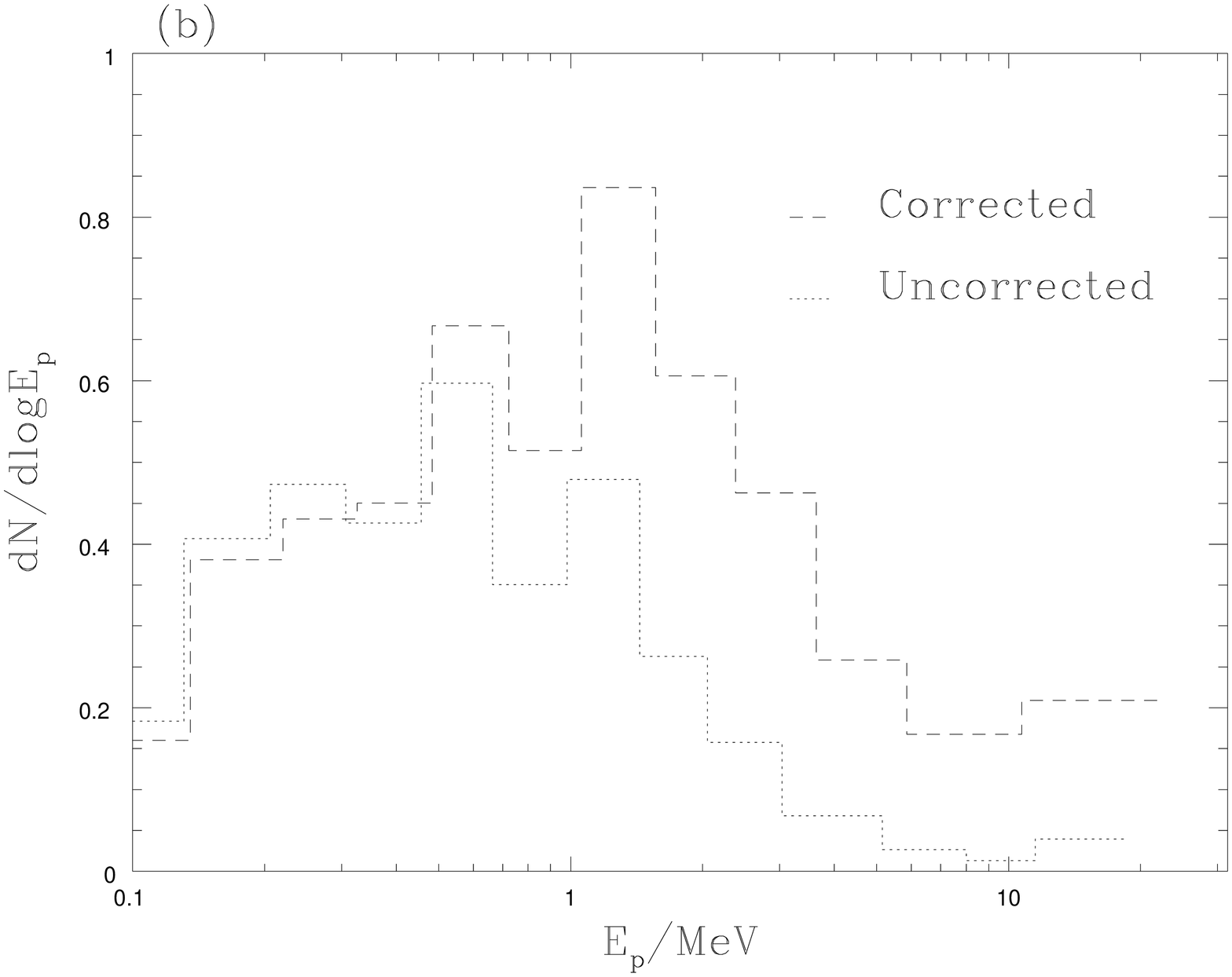,width=6.8cm,height=6cm,angle=270}}
     \label{(b)}
     \caption{Figure (a) shows the corrected and uncorrected cumulative distributions
(calculated from above and below) of $E_{p}$,  where the total number of gamma ray bursts
is normalized to 1.  Figure (b) displays the corrected
 and uncorrected differential distributions
of $E_{p}$, normalized to the low end of the distribution to
emphasize the relative difference between the number of bursts with $E_{p}$ above and
below 1MeV}  
     \end{figure}
     above 1MeV in the corrected distribution, not present in the uncorrected distribution. 


We test the hypothesis that the two distributions are from the same parent distribution,
    using the Kolmogorov-Smirnov (K-S) test (for the cumulative distributions) and $\chi ^{2}$ test
    (for the differential distributions).  The $\chi ^{2}$ test is performed dividing
    the uncorrected data into two and three bins of equal numbers of bursts.
    The probabilities that the above hypothesis is true are shown below for each test.

\begin{center}
\begin{tabular} {lccr} \hline \hline
Test                 &  Statistic         &    Probabilities \\ \hline
K-S                  &  $0.26$            &    $5\times 10^{-13}$  \\
$\chi ^{2}$ (2 bins, 216 bursts/bin) &  $42.7$            &    $6\times 10^{-11}$  \\
$\chi ^{2}$ (3 bins, 144 bursts/bin) &  $58.5$            &    $2\times 10^{-13}$  \\ \hline \hline
\end{tabular}
\end{center}
As shown elsewhere in these proceedings (Harris et al (11)), the corrected distribution agrees
well with the distribution of $E_{p}$ obtained from untriggered SMM data.

As a by product of this work, we are able to use the spectral parameters we obtained
      for each burst to correct the observed fluence (50-300keV) to the total fluence.  Figure 3
      shows the $Log(N)-Log(S)$ distributions for total fluence (essentially the
      distribution of $A$ in equation (1) ) of our entire sample, as
      well as the distribution for bursts with $E_{p}$ above 1MeV and $E_{p}$ below 400keV.
      \begin{figure}
      \centerline{\epsfig{file=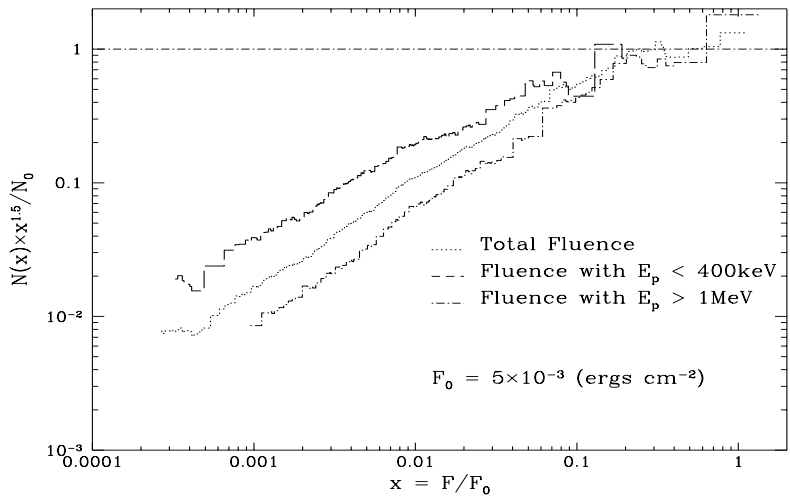,height=7cm,width=11cm,angle=270}}
      \caption{ Cumulative Distributions of the  $total$ 
      fluence for 433 GRBs, as well as the distributions
      for GRBs with $E_{p} < 400keV$ (172 GRBs) and $E_{p} > 1MeV$ (129 GRBs)
      .  Deviation from the dashed
      line indicates deviation from homogeneous, isotropic, static, and Euclidean geometry (HISE).}
      \end{figure}

      In the future, with more accurate values of the spectral parameters, we can not
      only refine the corrected distribution of $E_{p}$, but also correct the observed
      distribution of $\alpha$ and $\beta$; we can then determine correlations between these parameters 
      and fluence or flux, accounting for the double sided truncation.

\end{document}